\documentclass[epj,twocolumn]{webofc}
\usepackage[varg]{txfonts}   
\usepackage{epsfig}
\usepackage{amssymb,amsmath}
\usepackage{slashed}
\DeclareMathOperator{\BR}{BR}

\woctitle{Hadron Collider Physics symposium 2012}
\begin{document}
\title{Minimal dilaton model}
%
%

\author{Tomohiro Abe\inst{1}\fnsep\thanks{\email{tomohiro_abe@tsinghua.edu.cn}} \and
        Ryuichiro Kitano\inst{2}\fnsep\thanks{\email{kitano@tuhep.phys.tohoku.ac.jp}} \and
        Yasufumi Konishi\inst{3}\fnsep\thanks{\email{konishi@krishna.th.phy.saitama-u.ac.jp}} \and
        Kin-ya Oda\inst{4}\fnsep\thanks{\email{odakin@gauge.scphys.kyoto-u.ac.jp}} \and
        Joe Sato\inst{3}\fnsep\thanks{\email{joe@phy.saitama-u.ac.jp}} \and
        Shohei Sugiyama\inst{5}\fnsep\thanks{\email{shohei@icrr.u-tokyo.ac.jp}}
}

\institute{Institute of Modern Physics and Center for High Energy Physics, Tsinghua University, Beijing 100084, China
\and
     Department of Physics, Tohoku University, Sendai 980-8578, Japan
\and
	Department of Physics, Saitama University, Saitama 355-8570, Japan
\and
	Department of Physics, Kyoto University, Kyoto 606-8502, Japan
\and
	Institute for Cosmic Ray Research (ICRR), University of Tokyo, Kashiwa, Chiba 277-8582, Japan
}

\abstract{%
Both the ATLAS and CMS experiments at the LHC have reported the observation of the particle of mass around 125\,GeV which is consistent to the Standard Model (SM) Higgs boson, but with an excess of events beyond the SM expectation in the diphoton decay channel at each of them. There still remains room for a logical possibility that we are not seeing the SM Higgs but something else. Here we introduce the minimal dilaton model in which the LHC signals are explained by an extra singlet scalar of the mass around 125\,GeV that slightly mixes with the SM Higgs heavier than 600\,GeV. When this scalar has a vacuum expectation value well beyond the electroweak scale, it can be identified as a linearly realized version of a dilaton field. Though the current experimental constraints from the Higgs search disfavors such a region, the singlet scalar model itself still provides a viable alternative to the SM Higgs in interpreting its search results.
}
\maketitle
\section{Introduction}
The ATLAS~\cite{:2012gk} and CMS~\cite{:2012gu} experiments have reported the observation of the particle of mass 125\,GeV that is consistent with the Standard Model (SM) Higgs boson, but with an excess of events in the diphoton decay channel with the signal strength $1.80\pm0.30\pm0.7$ at ATLAS~\cite{ATLAS diphoton} and $1.6_{-0.6}^{+0.7}$ at CMS~\cite{CMS combined}. There still remain a logical possibility that the observed particle is not the SM Higgs but an extra boson beyond the SM. 

Here we report the minimal dilaton model~\cite{Abe:2012eu} in which the observed particle is identified as the SM singlet scalar field $S$ of mass around 125\,GeV that slightly mixes with the SM Higgs $H$ heavier than 600\,GeV. A vector-like top partner $T$ is introduced so that $S$ can couple to a pair of gluons or photons through the $T$ loop, just as in the case of the SM Higgs coupling to them via the top quark loop. The top partner $T$ at the same time cures the constraints from the electroweak precision data by its loop driving the Peskin-Takeuchi T-parameter positive and hence canceling the heavy SM Higgs contributions.

If the vacuum expectation value (vev) of the singlet $f:=\left\langle S\right\rangle$ is much larger than the electroweak scale $v\simeq246\,$GeV, then $S$ can be identified as a linear realization of a dilaton field  (hence the model's name) associated with an almost scale invariant sector whose strongly coupled interactions, broken at $f$, give rise to the top quark and the SM Higgs as composite particles, and the top partner~$T$ represents a fermion in the composite sector~\cite{Abe:2012eu}.

In literature~\cite{conventional dilaton}, the terminology ``dilaton model'' has been used for the one in which all the SM particles couple to the dilaton through the trace of the energy momentum tensor of the SM. This type of model implicitly assumes that all the SM particles are composite under the strong dynamics in the ultraviolet region. On the other hand, we take more conservative approach that the SM except for the top/Higgs sector is a spectator of the dynamics, and thus the dilaton $S$ couples to the $W$, $Z$ bosons and to light fermions only through the mixing with the Higgs fields $H$, while the couplings to the gluons and photons are generated only through the loops of the top quark and its partner. Due to these different origin of the couplings between two models, the production and decay properties are quite different. Indeed, we see that our model can give better fit to the LHC data compared to the SM Higgs boson, while the authors of Ref.~\cite{conventional dilaton} have reported that the above mentioned dilaton scenario is rather disfavored.

\section{Minimal dilaton model}
Our starting Lagrangian~\cite{Abe:2012eu} can be written in terms of the singlet scalar $S$, the SM Higgs $H$, the left handed quark doublet of the third generation $q_{3L}$, and the top partner $T$ as
\begin{align}
\mathcal L
	&=	\mathcal L_\text{SM}
		-{1\over2}\partial_\mu S\partial^\mu S
		-\tilde V(S,H)\nonumber\\
	&\quad
		-\overline T\left(\slashed{D}+yS\right)T
		-\left[y'\overline T(q_{3L}\cdot H)+\text{h.c.}\right],
\end{align}
where $\mathcal L_\text{SM}$ is the SM Lagrangian except for the Higgs potential and $\tilde V(S,H)$ is the  potential for the scalars
\begin{align}
\tilde V(S,H)
	&=	{m_S^2\over2}S^2+{\lambda_S\over 4!}S^4+{\kappa\over2}S^2\left|H\right|^2
		+m_H^2\left|H\right|^2+{\lambda_H\over4}\left|H\right|^4.
		\label{potential}
\end{align}
The detailed potential shape~\eqref{potential} is unimportant, as its couplings can be translated into the following two parameters relevant to the Higgs searches: the Higgs-singlet mixing~$\theta_H$ and the ratio of the vevs $\eta:=v/f$. The relation is shown in Appendix B in Ref.~\cite{Abe:2012eu}. We identify the lighter mass eigenstate to be the observed 125\,GeV boson and the heavier to be almost the SM Higgs with mass $\gtrsim 600$\,GeV. The vector-like top partner $T$ is set to be an $SU(2)_L$ singlet and to have a hypercharge $Y=2/3$ so that it can decay into the top quark through a small mixing $\theta_L\sim y'v/\sqrt{2}yf$. 

\begin{figure}
\begin{center}
\includegraphics[width=.3\textwidth]{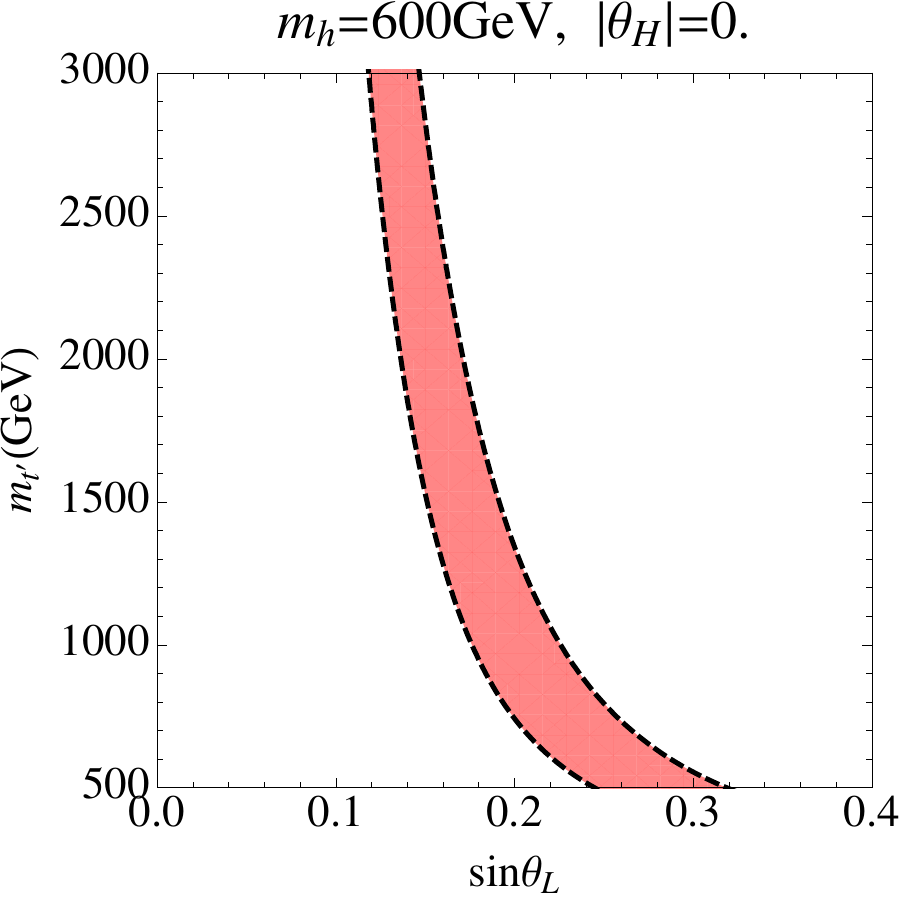}\bigskip\\
\includegraphics[width=.3\textwidth]{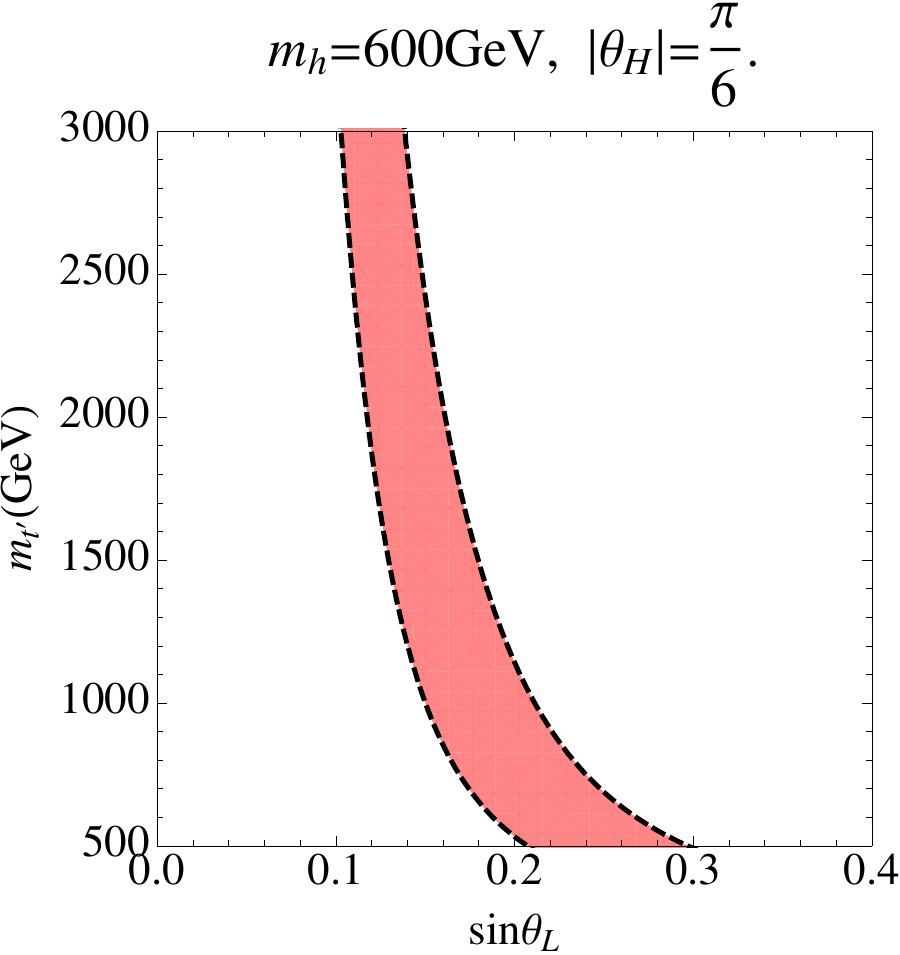}\bigskip\\
\includegraphics[width=.3\textwidth]{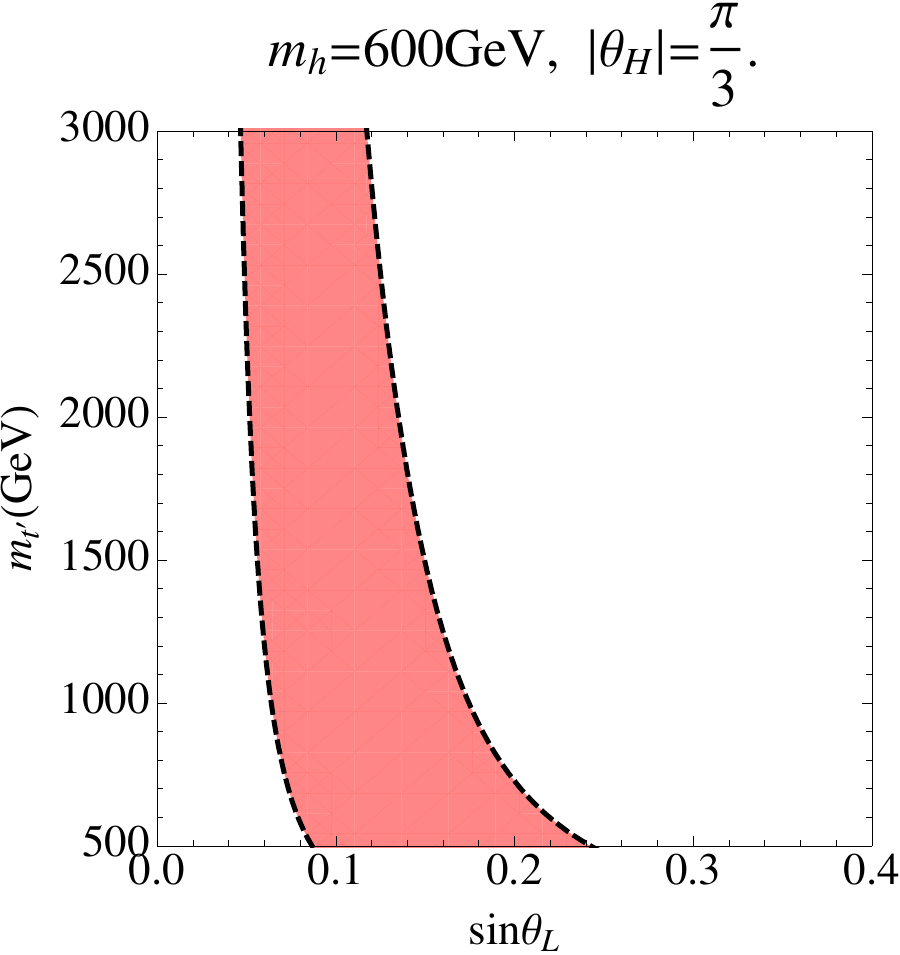}
\caption{Allowed region plot by the Peskin-Takeuchi S,T parameters at the 95\% CL in the plane of top-partner mass $m_{t'}$ vs top-top partner mixing $\theta_L$; The SM Higgs mass is taken to be $m_h=600$\,GeV and the Higgs-singlet mixing angle is chosen as $\theta_H=0$, $\pi/6$, and $\pi/3$~\cite{Abe:2012eu}.}\label{electroweak constraints}
\end{center}
\end{figure}

When the SM Higgs mass is heavy, the Peskin-Takeuchi S,T parameters severely constrain the model. In the SM, the Higgs mass heavier than 185\,GeV is ruled out at the 95\% CL~\cite{ALEPH:2010aa}. In the minimal dilaton model, the $T$ loop gives positive contribution to the Peskin-Takeuchi T-parameter and cures the model from the heavy Higgs effects. In Fig.~\ref{electroweak constraints}, we show the constraints from the S,T parameters on the top-partner mass $m_{t'}$ and the mixing $\theta_L$ between the top and top partner, at the 95\% CL for the SM Higgs mass $m_h=600$\,GeV and the Higgs-singlet mixing angle $\theta_H=0$, $\pi/6$, and $\pi/3$~\cite{Abe:2012eu}.

\section{Higgs signal at LHC}
Let us show our prediction for the Higgs searches at the LHC. In Ref.~\cite{conventional dilaton}, the parameter $c_X$ is defined as the ratio of the coupling of $S$ to $XX$ (or $X\bar X$), in the amplitude, to that of the SM Higgs at 125\,GeV. The minimal dilaton model gives~\cite{Abe:2012eu}
\begin{align}
c_V	=	c_F	
	&=	\sin\theta_H,\nonumber\\
c_t	&=	\cos^2\theta_L\sin\theta_H+\eta\sin^2\theta_L\cos\theta_H,\nonumber\\
c_g	&=	\eta\cos\theta_H+\sin\theta_H,\nonumber\\
c_\gamma
	&=	\eta A_{t'}\cos\theta_H+A_\text{SM}\sin\theta_H,\nonumber\\
c_\text{inv}
	&=	0,
\end{align}
where $A_{t'}\simeq16/9$, $A_\text{SM}\simeq-6.5$, the subscript ``$F$'' stands for all the SM fermions except the top quark, and ``$V$'' for $W$ or $Z$.

\begin{figure}
\begin{center}
\includegraphics[width=.3\textwidth]{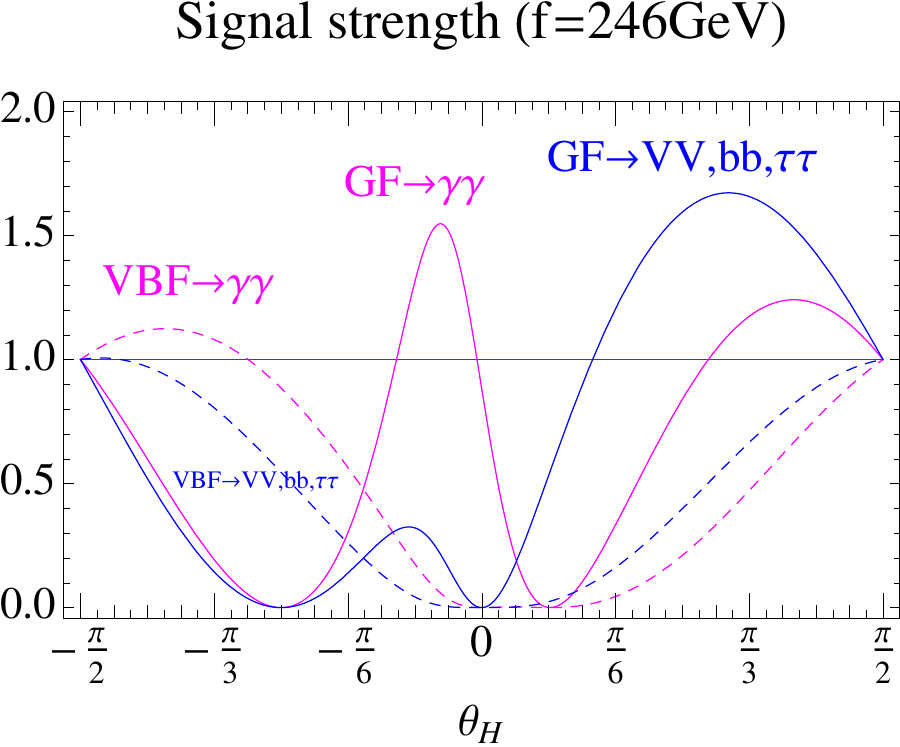}\bigskip\\
\includegraphics[width=.3\textwidth]{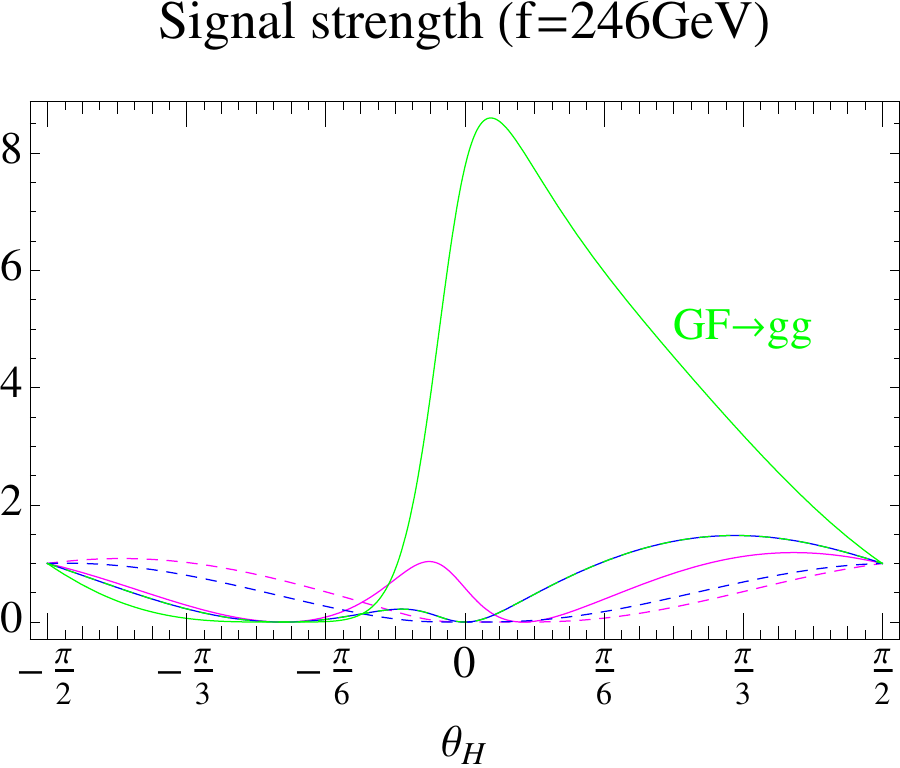}
\caption{Upper: Signal strength for $f=246$\,GeV ($\eta=1$). ``GF'' denotes the case that $S$ is produced by the gluon fusion process, while ``VBF'' by the vector boson fusion, Higgs-strahlung, or $ttH$ process. Lower: The same for the Higgs to digluon process; Shown for comparison though hardly observable at the LHC.}\label{signal strength}
\end{center}
\end{figure}

In Fig.~\ref{signal strength}, we plot the resultant signal strength, which is the ratio of the model cross section to the corresponding SM Higgs one at 125\,GeV, for the singlet vev $f=246$\,GeV ($\eta=1$). We see that the diphoton signal can be enhanced in the dilatonic region $\theta_H\sim 0$, whereas other processes are suppressed.

The minimal dilaton model predicts different production cross sections
between GF and VBF/VH/ttH processes. In $H\to\gamma\gamma$ search,
composition of these production channels differs category by category
and are summarized in Table 2 in Ref.~\cite{:2012gu} for CMS and in
Table~6 in Ref.~\cite{ATLAS_diphoton} for ATLAS. 
We define $\varepsilon^i_X$ as the proportion of the production process
$X$ within a category $i$.
Note that $\sum_X\varepsilon^i_X=1$ by definition for each category
$i$, where a summation over $X$ is always understood as for all the relevant production channels: GF, VBF, VH, and ttH.
GF is the dominant production process
and satisfies $\varepsilon^i_\text{GF}\lesssim 90\%$
in production processes other than dijet category. 
In the dijet category, the dominant production process is VBF,
and $\varepsilon_\text{VBF}\lesssim 70\%$.

When acceptance of a production channel $X$ for a category $i$ is $a^i_X$, the estimated value of a signal fraction under the given set of cuts $i$ becomes
\begin{align}
\varepsilon^i_X	&=	{a^i_X\sigma^\text{SM}_X\over\sum_Ya^i_Y\sigma^\text{SM}_Y},
	\label{varepsilon_eq}
\end{align}
where $\sigma^\text{SM}_X$ is the Higgs production cross section in the SM through the channel $X$.
Given $\{\varepsilon^i_X\}$, we can compute the signal strength under the imposed cuts for each category $i$
\begin{align}
\hat\mu^i(h\to \gamma\gamma)
	&=	{\sum_Xa^i_X\,\sigma_X\over
			\sum_Y a^i_Y\,\sigma_Y^\text{SM}}\,{\BR(s\to \gamma\gamma)\over \BR(h\to\gamma\gamma)_\text{SM}}
			\nonumber\\
	&=	\sum_X\varepsilon^i_X \left|c_X\right|^2
				\,{\left|c_\gamma\right|^2\over R(s\to\text{all})},
				\label{signal_strength}
\end{align}
where $R(s\to\text{all}):=0.913\left|c_V\right|^2+0.085\left|c_g\right|^2+0.002\left|c_\gamma\right|^2$.
We have assumed that the acceptance $a^i_X$ under the category $i$ does not change from that of the SM for each production channel $X$. 

For the $ZZ\to4l$ and $WW\to l\nu l\nu$ decay channels, we assume that all the signals are coming from GF and, hence, we approximate
\begin{align}
\hat\mu(s\to VV)
	&=	\left|c_g\right|^2\,{\left|c_V\right|^2\over R(s\to\text{all})}
\end{align}
for $VV=WW$ and $ZZ$. 

As all the signal strengths are obtained, we perform a chi-square test with the Gaussian approximation for all the errors
\begin{align}
\chi^2
	&=	\sum_i\left(\hat\mu_i-\mu_i\over\sigma_i\right)^2,
\end{align}
where summation over $i$ is for all the diphoton categories as well as the $WW$ and $ZZ$ channels.
Of course this is a naive fit without taking into account the off-diagonal elements of the correlation matrix for various categories. For example, this type of naive weighting does not reproduce the central value nor the error of the diphoton signal strength. Although we are aware of this fact, this is the best we can do with the current data made public. The result should be regarded as an illustration at best. 

\begin{figure}
\begin{center}
\includegraphics[width=.3\textwidth]{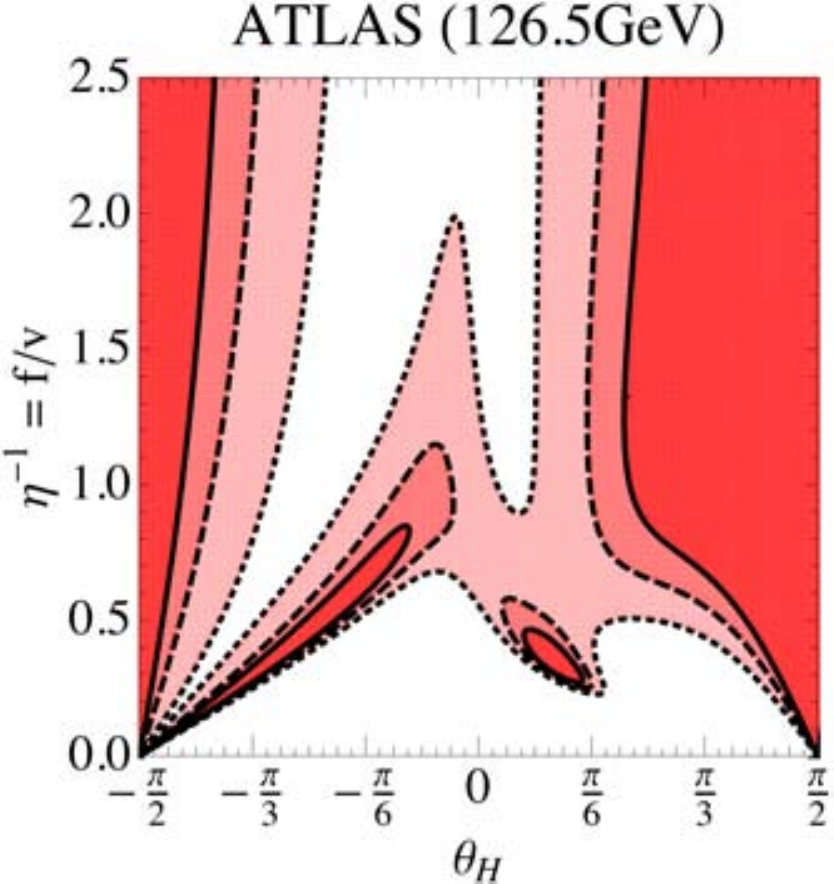}\bigskip\\
\includegraphics[width=.3\textwidth]{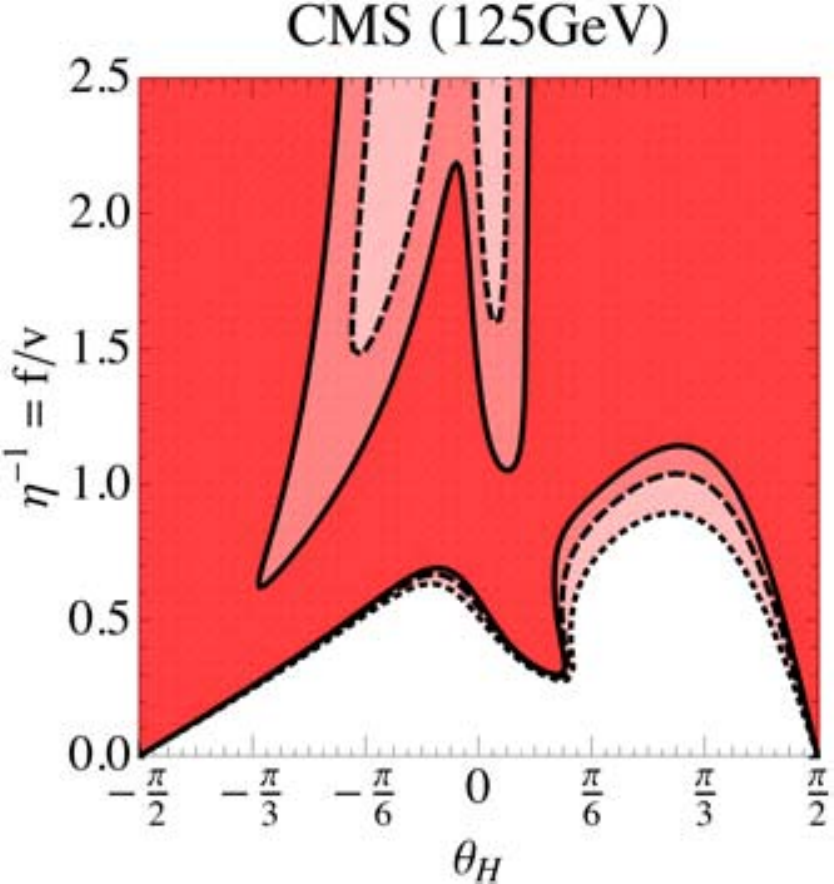} 
\caption{Favored regions within 90, 95 and 99\% confidence intervals, enclosed by solid, dashed, and dotted lines, respectively. Density (area) of favored region decreases (increases) in according order~\cite{Abe:2012eu}.}\label{Higgs constraints}
\end{center}
\end{figure}

\begin{figure}
\begin{center}
\includegraphics[width=.3\textwidth]{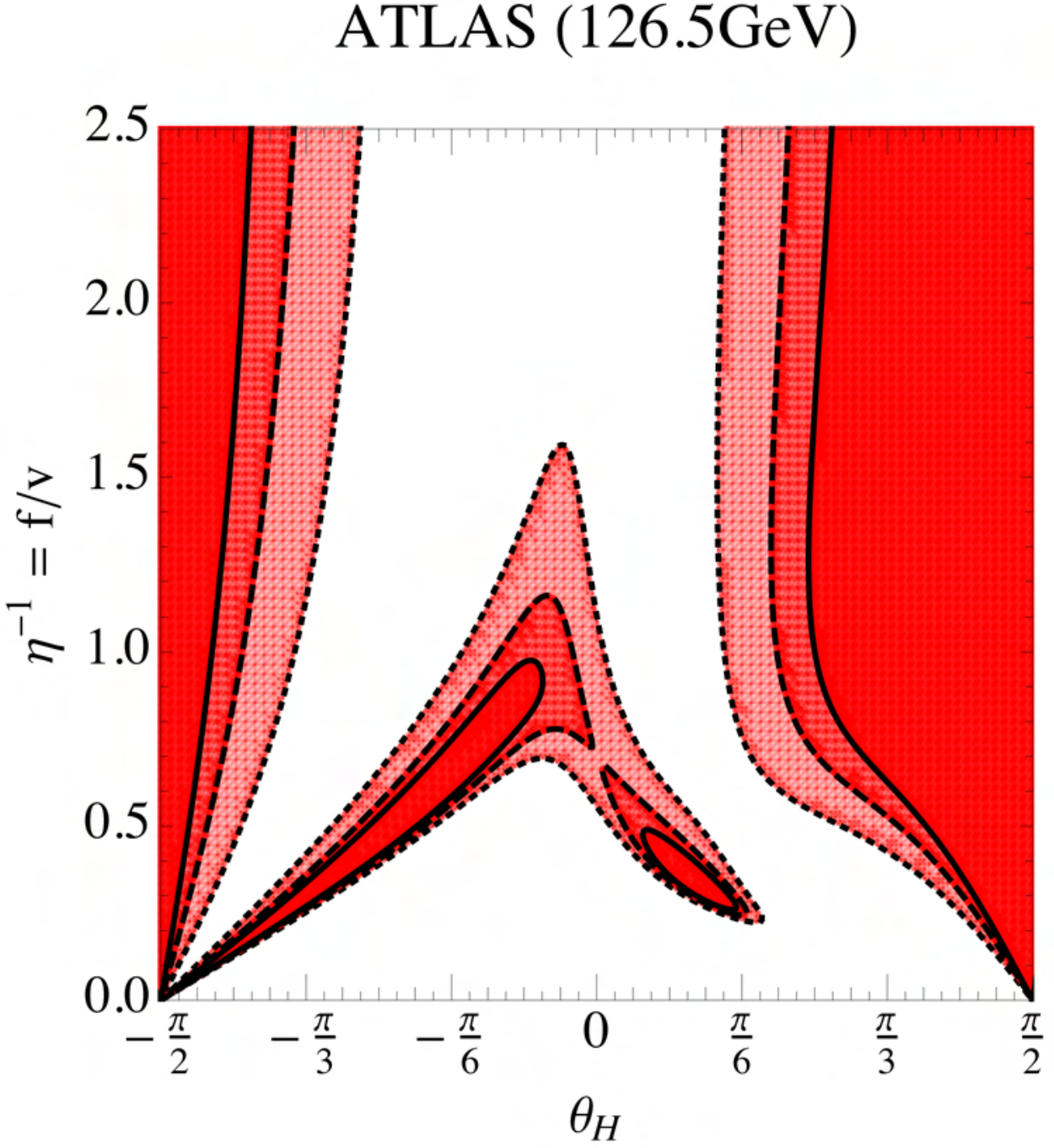}\bigskip\\
\includegraphics[width=.3\textwidth]{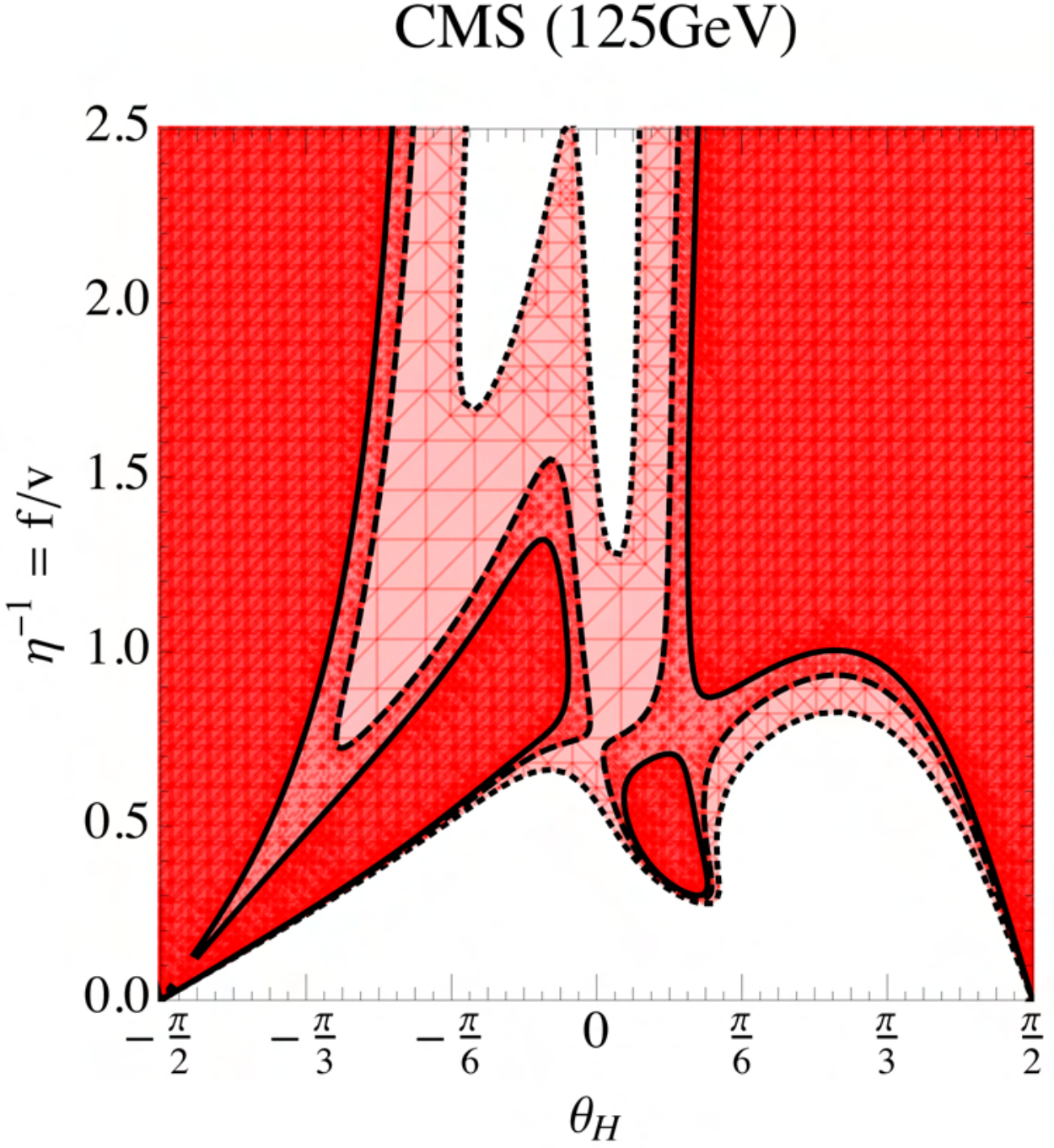}
\caption{The same as Fig.~\ref{Higgs constraints}, including the data presented at HCP2012 and after.}\label{Higgs constraints HCP}
\end{center}
\end{figure}

Keeping the above caution in mind, we show the constraints from the data from the LHC Higgs searches in Fig~\ref{Higgs constraints}. For comparison, we also show in Fig.~\ref{Higgs constraints HCP} the same plot including more recent data presented at HCP2012 and after~\cite{HCP and after}. We see that the dilaton like region $\eta^{-1}\gg1$, $\left|\theta_H\right|\ll1$ is disfavored by the Higgs data alone, but there still remains phenomenologically viable region at $\left|\theta_H\right|\lesssim\pi/6$ and $\eta^{-1}\lesssim1.5$.

\section{Summary}
We have examined the possibility that the 125\,GeV boson observed at the LHC is not the SM Higgs but an extra singlet scalar $S$. If the vev well exceeds the electroweak scale, $f=\left\langle S\right\rangle\gg 246$\,GeV, then $S$ can be regarded as a linearly realized dilaton that is associated with the quasiconformal dynamics making up the top and Higgs particles as composite ones. However, such a parameter region is disfavored by the current Higgs data. Though such a linear dilaton interpretation is marginally excluded, this model with a singlet scalar and a vector-like top partner still provides a phenomenologically viable alternative to the SM Higgs to fit the current LHC data.

\section*{Acknowledgments}
This work is supported in part by the Grants-in-Aid for Scientific Research No.~23740165 (R.~K.), No.~23104009, No.~20244028, No.~23740192 (K.~O.), and No.~24340044 (J.~S.) of JSPS.


  

%
%

\end{document}